\documentclass[preprint,12pt]{elsarticle}

\usepackage[T1]{fontenc}
\usepackage[utf8]{inputenc}
\usepackage{amsmath,amssymb,amsfonts}
\usepackage{bm}
\usepackage{graphicx}
\usepackage{microtype}
\usepackage{float}

\newcommand{\Sg}{\mathcal S_g}
\newcommand{\Om}{\mathcal O_m}

\newcommand{\Pbar}{\overline P}

\begin{document}
\begin{frontmatter}

\title{Chromoelectric and chromomagnetic quarkonium couplings to scalar and
spin-two nucleon structure}

\author[bitp]{Arkadiy I. Syamtomov\corref{cor1}}
\cortext[cor1]{Corresponding author}
\ead{arkadys@bitp.kyiv.ua}
\address[bitp]{Bogolyubov Institute for Theoretical Physics, National Academy
of Sciences of Ukraine, Kyiv, Ukraine}

\begin{abstract}
The chromoelectric and chromomagnetic polarizabilities
of compact heavy quarkonium weight the scalar and
spin-two structure of the nucleon differently.
We derive the corresponding finite-momentum-transfer
elastic response kernels, retaining the physical
quarkonium velocity and nucleon spin dependence.
Their correlated uncertainties are determined using
lattice quark and gluon gravitational form factors.
For $\Upsilon(1S)N$ at $100\,\mathrm{MeV}$ above
threshold and $Q^2=0.20\,\mathrm{GeV}^2$, the electric
non-spin-flip kernel is
$1.316\pm0.100_{\rm GFF}\,\mathrm{GeV}$.
The electric and magnetic kernels exhibit distinct
momentum-transfer and spin dependence.
These kernels describe the nucleon response to
the local operators. Their contributions to elastic
scattering are weighted by the independent quarkonium
polarizabilities.
\end{abstract}

\end{frontmatter}

\section{Introduction}
\label{sec:intro}

The interaction of compact heavy quarkonium with hadronic matter offers a way
to probe the gluonic structure of the nucleon. Because the heavy quark and
antiquark form a small color-singlet system, their interaction with soft
gluonic fields can be described through chromoelectric and chromomagnetic
polarizabilities. This connection was established in the short-distance
approach of Peskin \cite{Peskin:1979,Peskin:1979II} and developed for quarkonium interactions with nuclei by
Luke, Manohar, and Savage (LMS) \cite{LukeManoharSavage:1992}.
The low-energy quarkonium--nucleon interaction has also been studied in a
hadronic effective field theory by Tarr\'us Castell\`a and Krein
\cite{TarrusCastellaKrein:2018}. The electric and magnetic interactions
are governed by different combinations of
scalar and traceless spin-two gluonic operators, reflecting distinct responses
of quarkonium to the fields inside the nucleon.

The spin-two gluon operator contributes to the traceless
QCD energy-momentum tensor (EMT), while the scalar gluon
operator is related to the trace of the total EMT and
the quark-mass contribution through the trace anomaly \cite{CollinsDuncanJoglekar:1977}.
The corresponding nucleon matrix elements describe
distinct aspects of nucleon structure, including the
distribution of energy and momentum between quarks and
gluons. At finite momentum transfer, the EMT matrix
elements are parametrized by gravitational form factors.
Hoodbhoy examined off-forward gluonic matrix elements
in quarkonium scattering \cite{Hoodbhoy:1999}, while
the scalar and spin-two components of the chromoelectric
response have also been studied in a light-front
formulation \cite{Syamtomov:2026PLB}.

The chromoelectric and chromomagnetic interactions
couple to different linear combinations of the
scalar and spin-two nucleon matrix elements.
Their relative weights depend on the momentum
transfer, the quarkonium four-velocity, and the
nucleon spin transition. A quantitative
determination therefore requires both the quark and gluon gravitational form
factors and their correlations. The lattice calculation of Hackett, Pefkou, and
Shanahan provides these form factors
\cite{HackettPefkouShanahan:2024}, while QCD
evolution and heavy-flavor matching connect their renormalization scale to that
of the quarkonium interaction \cite{HillSolon:2015}.

We determine the scalar and spin-two contributions to the finite-transfer
electric and magnetic quarkonium--nucleon response kernels, retaining the
physical quarkonium velocity and nucleon spin dependence. Correlated lattice
gravitational form factors give their momentum-transfer dependence and
uncertainties. The resulting interference patterns distinguish how the two
quarkonium polarizabilities probe nucleon structure.

\section{Quarkonium couplings and QCD matching}
\label{sec:matching}

The interaction of compact $S$-wave quarkonium with
soft gluonic fields is characterized by its chromoelectric
and chromomagnetic polarizabilities. 
The leading electric response arises from E1
transitions through color-octet intermediate states
\cite{Brambilla:2016},
while the magnetic response involves sequential M1
transitions and a local contribution from the
gauge-covariant heavy-quark kinetic interaction
\cite{SyamtomovBOEFT:2026}. At leading order in the local two-gluon expansion,
the interaction can be expressed in terms of the scalar and spin-two gluonic operators as
\begin{equation}
\begin{aligned}
\footnotemark
\mathcal L_{\rm pol}^{[4]}
={}&\frac12\Phi_v^\dagger\Phi_v
\left[
C_2^\Phi(\mu_\Phi)\,
v_\mu v_\nu
O_g^{(2)\mu\nu}(\mu_\Phi)
\right.\\
&\left.
+C_S^\Phi(\mu_\Phi)\,
\mathcal S_g(\mu_\Phi)
\right],\\
C_2^\Phi
={}&\alpha_E^\Phi+\alpha_B^\Phi,
\qquad
C_S^\Phi=\alpha_B^\Phi-\alpha_E^\Phi.
\end{aligned}
\label{eq:matching}
\end{equation}
\footnotetext{The label $[4]$ denotes the four-active-flavor
QCD theory at the quarkonium matching scale.
The corresponding three-flavor interaction
is obtained by integrating out charm.}
Here the scalar and traceless spin-two gluonic
operators are defined by
\begin{equation}
\begin{aligned}
\mathcal S_{g}
&=\frac14[G_{\alpha\beta}^a
G^{a\alpha\beta}],\\
O_{g}^{(2)\mu\nu}
&=\left[
-G^{a\mu\lambda}G^{a\nu}{}_\lambda
+\frac14g^{\mu\nu}
G_{\alpha\beta}^aG^{a\alpha\beta}
\right].
\end{aligned}
\label{eq:gluon-operators}
\end{equation}
The brackets denote renormalized operators
in the $\overline{\mathrm{MS}}$ scheme.
The gluon spin-two operator mixes under
renormalization with the flavor-singlet quark
spin-two operator, while their sum is
renormalization-group invariant.

In the weak-coupling regime
$m_Qv_Q\gg\Lambda_{\mathrm{QCD}}$,
the Coulombic estimates
$r_\Phi\sim(m_Qv_Q)^{-1}$ and
$\Delta E\sim m_Qv_Q^2$
give the characteristic spatial and
excitation-energy scales.
The sequential M1 contribution and the quadratic
kinetic magnetic term are both of relative order
$v_Q^2$ compared with the leading E1--E1 response.
The former involves intermediate-state propagation,
while the latter is local. This counting assumes
excitation gaps of Coulombic order and chromofields
varying slowly across the quarkonium state.
The coefficient $\alpha_B^\Phi$ parametrizes
the spin-independent local magnetic response
of quarkonium, accounting for both mechanisms.

The QCD trace identity \cite{CollinsDuncanJoglekar:1977}
relates the scalar gluon interaction to the total
EMT trace $\Theta=T^\mu{}_\mu$ and the quark-mass
operator $\Om=\sum_qm_q[\bar q q]$ as follows:
\begin{equation}
\begin{aligned}
C_S^\Phi\Sg
={}&C_\Theta^\Phi\Theta
+C_m^{\Phi,\mathrm{inv}}\Om,\\
C_\Theta^\Phi
={}&C_S^\Phi\frac{g}{2\beta(g)},\\
C_m^{\Phi,\mathrm{inv}}
={}&-(1+\gamma_m)C_\Theta^\Phi.
\end{aligned}
\label{eq:scalar-map}
\end{equation}
These relations define the scalar coefficients at
the quarkonium matching scale $\mu_\Phi$.
The corresponding three-flavor coefficients are
obtained through QCD evolution and heavy-quark
decoupling, while the spin-two coefficients evolve
through quark-gluon mixing \cite{HillSolon:2015}.

The matching then could be used to connect the interaction at
$\mu_\Phi=3\,\mathrm{GeV}$ to the three-flavor
nucleon form factors at $2\,\mathrm{GeV}$.
We choose the charm matching scale
$\mu_c=\overline m_c(\mu_c)=1.27\,\mathrm{GeV}$. Then the NLO scalar relation is, according to 
\cite{HillSolon:2015}
\begin{equation}
\begin{aligned}
m_c[\bar c c]
\longrightarrow{}&
-\frac{\alpha_s^{(4)}(\mu_c)}{3\pi}
\left[
1+\frac{11\alpha_s^{(4)}(\mu_c)}{4\pi}
\right]\mathcal S_g^{[3]}
\\
&+O(\alpha_s^3,m_c^{-2}).
\end{aligned}
\label{eq:charm-NLO}
\end{equation}
The charm contribution enters both scalar coefficients through the three-flavor trace identity.
The resulting scalar-coefficient transformation is given in \ref{app:scalar-matching}.

With two-loop running of coupling constant $\alpha_s$, the one-loop
mass anomalous dimension, and NLO scalar matching,
the coefficient evolution gives
\begin{equation}
\begin{aligned}
\frac{C_{\Theta,3}}{C_S^\Phi}
&=-2.380,
&
\frac{C_{m,3}^{\mathrm{inv}}}{C_S^\Phi}
&=2.746,
\\
\frac{C_{q,3}^{(2)}}{C_2^\Phi}
&=0.06164,
&
\frac{C_{g,3}^{(2)}}{C_2^\Phi}
&=0.94048.
\end{aligned}
\label{eq:matched-coefficients}
\end{equation}
In the following, the flavor label is omitted
from the matched three-flavor coefficients.
The spin-two coefficients are evolved at
leading-logarithmic accuracy with tree-level
charm-threshold matching. These values correspond
to $\alpha_{s}^{(4)}(3\,\mathrm{GeV})=0.255$
and enter the nucleon amplitudes, which we discuss in the next section.

\section{Elastic amplitude and nucleon form factors} \label{sec:nucleon}

The local quarkonium interaction in
Eq.~\eqref{eq:matching} probes the scalar and
traceless spin-two nucleon matrix elements.
For initial and final nucleon four-momenta
$P$ and $P'$, we introduce the momentum transfer
$\Delta=P'-P$, its invariant square $t=\Delta^2$,
and the average nucleon four-momentum
$\Pbar=(P'+P)/2$. Then we can write the scalar matrix elements as
\begin{equation}
\begin{aligned}
\langle P',s'|\Theta|P,s\rangle
  &=\bar u'u\,\Theta_N(t),\\
\langle P',s'|\Om|P,s\rangle
  &=\bar u'u\,\sigma_N(t).
\end{aligned}
\label{eq:scalar-ffs}
\end{equation}
Here $\bar u'u=\bar u(P',s')u(P,s)$, with
$\bar u(P,s)u(P,s)=2M$. The form factors
$\Theta_N(t)$ and $\sigma_N(t)$ describe the
total QCD trace and the quark-mass contribution,
respectively. At zero momentum transfer, the
light- and strange-quark scalar contributions
define the nucleon sigma terms,
\begin{equation}
\begin{aligned}
\sigma_{\pi N}
&=\frac{\hat m}{2M}
  \langle P,s|\bar u u+\bar d d|P,s\rangle,\\
\sigma_s
&=\frac{m_s}{2M}
  \langle P,s|\bar s s|P,s\rangle,
\end{aligned}
\label{eq:sigma-terms}
\end{equation}
where $\hat m=(m_u+m_d)/2$. In the
isospin-symmetric limit we have
$\sigma_N(0)=\sigma_{\pi N}+\sigma_s$.

On the other hand side, the traceless quark and gluon EMT matrix elements could be conveniently 
parametrized (see, e.g. \cite{Ji:1997b}) by the gravitational form factors
$A_i$, $B_i$, and $D_i$. If we rewrite the operator matrix element, using the 
Dirac vertex function $\Gamma_i^{\mu\nu}$,
\begin{equation}
\begin{aligned}
\langle P',s'|O_i^{(2)\mu\nu}(\mu)|P,s\rangle
={}&\bar u(P',s')
\left[
\Gamma_i^{\mu\nu}
-\frac14 g^{\mu\nu}
\Gamma_{i,\rho}{}^\rho
\right]u(P,s),
\end{aligned}
\label{eq:spin2-matrix}
\end{equation}
the latter could be expressed through $A_i$, $B_i$, and $D_i$ as
\begin{equation}
\begin{aligned}
\Gamma_i^{\mu\nu}(P',P;\mu)
={}&A_i(t,\mu)\gamma^{(\mu}\Pbar^{\nu)}
\\
&+B_i(t,\mu)
\frac{\Pbar^{(\mu}i\sigma^{\nu)\rho}
\Delta_\rho}{2M}
\\
&+D_i(t,\mu)
\frac{\Delta^\mu\Delta^\nu-g^{\mu\nu}t}{4M}.
\end{aligned}
\label{eq:emttensor}
\end{equation}
Here $i=q,g$, $B_i=2J_i-A_i$, and the upper parentheses
denote symmetrization with a factor $1/2$.
In Eq.~\eqref{eq:emttensor}, terms proportional to $g^{\mu\nu}$ are removed by
the traceless projection.

The velocity-dependent interaction term in
Eq.~\eqref{eq:matching} probes the projection of
the nucleon spin-two EMT operator along the quarkonium four-velocity. Therefore, this introduces an
explicit dependence on the collision energy in addition to the momentum
transfer. The gravitational form factors above enter the different nucleon spin
transitions through their Lorentz-Dirac structures in Eq.~\eqref{eq:emttensor}. Let's denote the normalized spin-two
matrix element by
\begin{equation}
\mathcal F_{i,s's}^{(2)}(v,t,\mu)
=
\frac{v_\mu v_\nu}{2M}
\langle P',s'|
O_i^{(2)\mu\nu}(\mu)
|P,s\rangle.
\label{eq:F2-general}
\end{equation}
Combining the scalar and spin-two contributions together gives the leading local elastic
amplitude,
\begin{equation}
\begin{aligned}
\mathcal A_{\Phi N,s's}(v,t)
={}&\frac{\bar u'u}{2M}
\left[
C_\Theta\Theta_N(t)
+C_m^{\mathrm{inv}}\sigma_N(t)
\right]
\\
&+\sum_{i=q,g}C_i^{(2)}(\mu)
\mathcal F_{i,s's}^{(2)}(v,t,\mu).
\end{aligned}
\label{eq:master}
\end{equation}

If we denote the elastic initial and final quarkonium momenta as $k$ and $k'$, the
symmetric physical four-velocity then is
\begin{equation}
\begin{aligned}
v_\Phi^\mu
&=\frac{(k+k')^\mu/2}
{\sqrt{M_\Phi^2-t/4}},
&
v_\Phi\cdot\Delta&=0,
\\
v_\Phi\cdot\Pbar
&=\frac{s-u}
{4\sqrt{M_\Phi^2-t/4}},
&
s&=(k+P)^2,\quad u=(k-P')^2.
\end{aligned}
\label{eq:vphys}
\end{equation}
This velocity specifies the symmetric kinematic projection of the local
interaction. The analogous projection along the average nucleon
four-momentum, evaluated in the Breit frame, is given in \ref{app:breit}.

Using the matched coefficients (Eq.~\eqref{eq:matched-coefficients})
from Sec.~\ref{sec:matching}, let's separate the scalar
and spin-two nucleon responses into the kernels
\begin{equation}
\begin{aligned}
K_{S,s's}(s,t)
={}&\frac{\bar u'u}{2M}
\left[
\frac{C_\Theta^\Phi}{C_S^\Phi}\Theta_N(t)
+\frac{C_m^{\Phi,\mathrm{inv}}}{C_S^\Phi}
\sigma_N(t)
\right],
\\
K_{2,s's}(s,t)
={}&\sum_{i=q,g}
\frac{C_i^{(2),\Phi}(\mu)}{C_2^\Phi}
\mathcal F_{i,s's}^{(2)}(v_\Phi,t,\mu),
\\
K_{E,s's}(s,t)
={}&K_{2,s's}(s,t)-K_{S,s's}(s,t),
\\
K_{B,s's}(s,t)
={}&K_{2,s's}(s,t)+K_{S,s's}(s,t).
\end{aligned}
\label{eq:kernels}
\end{equation}
Note, that the chromoelectric and chromomagnetic responses contain the same spin-two contribution but opposite scalar contributions.
Using Eq.~\eqref{eq:kernels} the physical amplitude can therefore be written as
\begin{equation}
\mathcal A_{\Phi N,s's}(s,t)
=
\alpha_E^\Phi K_{E,s's}(s,t)
+\alpha_B^\Phi K_{B,s's}(s,t).
\label{eq:electric-magnetic}
\end{equation}
The kernels $K_E$ and $K_B$ in Eq.~\eqref{eq:kernels}  are fully determined by the matched QCD operators
and their nucleon matrix elements. The coefficients $\alpha_E^\Phi$ and
$\alpha_B^\Phi$, instead, describe the separate chromofield responses of the
quarkonium state. Their values are therefore required to determine the relative
contributions to the elastic amplitude.

\section{Physical quarkonium response from nucleon form factors}
\label{sec:quantitative}

Let's now evaluate the electric and magnetic responses using lattice quark and
gluon gravitational form factors (GFFs) and the nucleon scalar form factors.

We will use the quark and gluon GFFs, calculated on the lattice by Hackett,
Pefkou, and Shanahan~\cite{HackettPefkouShanahan:2024} in the three-flavor
$\overline{\mathrm{MS}}$ scheme at $2\,\mathrm{GeV}$.
The 400 correlated lattice values determine $A_i(t)$, $J_i(t)$,
and $D_i(t)$, with $B_i=2J_i-A_i$. 

Conservation of the total energy-momentum tensor implies the following relation for the trace
form factor:
\begin{equation}
\Theta_N(t)=M A_{\rm tot}(t)
+\frac{t}{4M}
\left[B_{\rm tot}(t)-3D_{\rm tot}(t)\right].
\label{eq:trace-GFF-fit}
\end{equation}
Here and below, the subscript ``tot'' denotes the
sum of the quark and gluon contributions to the
corresponding gravitational form factor,
$F_{\rm tot}(t)=\sum_{q=u,d,s}F_q(t)+F_g(t)$,
for $F=A,B,D,J$. 
We determine their momentum-transfer dependence
using a $z$ expansion, explicitly incorporating
the momentum and angular-momentum sum rules
$A_{\rm tot}(0)=1$ and $J_{\rm tot}(0)=1/2$.
The cubic expansion in our computation gives
$\chi^2/\mathrm{dof}=264.9/354$ and provides
the central GFF parametrization.

The scalar quark-mass form factor is determined, using the definitions ~\eqref{eq:sigma-terms}, from
$\sigma_{\pi N}(0)=43.7(3.6)\,\mathrm{MeV}$ and $\sigma_s(0)=28.6(9.3)\,\mathrm{MeV}$
\cite{Agadjanov:2023}, and extended to finite $t$ with $t$-dependent profile
\begin{equation}
\sigma_N(t)=\sum_{a=\ell,s}
\frac{\sigma_a(0)}
{\left[1-t\langle r_{S,a}^2\rangle/12\right]^2}.
\label{eq:scalar-shape}
\end{equation}
For $\sigma_{\pi N}=45\,\mathrm{MeV}$, the NLO+N2LO
dispersive analysis of Alarc\'on and Weiss
\cite{AlarconWeiss:2017} gives the interval
$1.34$--$1.49\,\mathrm{fm}^2$ for the light-quark
scalar squared radius. We adopt its midpoint,
$1.415\,\mathrm{fm}^2$, as the central value.
For the strange-quark contribution, we assume
a central squared radius of $0.55\,\mathrm{fm}^2$
and examine its influence below.

At $100\,\mathrm{MeV}$ above the $\Upsilon(1S)N$
threshold and $Q^2=0.20\,\mathrm{GeV}^2$, we obtain
the non-spin-flip response kernels
$K_E^{\uparrow\uparrow}
=1.3163\pm0.1003_{\rm GFF}\,\mathrm{GeV}$
and
$K_B^{\uparrow\uparrow}
=-0.7310\pm0.0930_{\rm GFF}\,\mathrm{GeV}$.
The opposite signs reflect the relative magnitudes
of the scalar and spin-two nucleon contributions,
which reinforce each other in the electric response
and partially cancel in the magnetic response.
Table~\ref{tab:physical} presents the spin-resolved
kernels and their dependence on the GFF
parametrization types. At these kinematics, their separate contributions are
\begin{equation}
\begin{aligned}
K_2^{\uparrow\uparrow}
  &=0.2927\,\mathrm{GeV},\\
K_S^{\uparrow\uparrow}
  &=-1.0237\,\mathrm{GeV}.
\end{aligned}
\label{eq:scalar-spin2-decomposition}
\end{equation}
The scalar contribution is approximately $3.5$
times the spin-two contribution in magnitude.

\begin{table}[t]
\centering
\small
\begin{tabular}{lrr}
\hline
Nucleon spin and form-factor shape & $K_E$ (GeV) & $K_B$ (GeV) \\
\hline
Non-spin-flip, cubic $z$
  & $1.3163\pm0.1003$ & $-0.7310\pm0.0930$ \\
Spin-flip, cubic $z$
  & $-0.0177\pm0.0044$ & $0.0688\pm0.0045$ \\
Non-spin-flip, quadratic $z$
  & $1.3233$ & $-0.7380$ \\
Non-spin-flip, quartic $z$
  & $1.3307$ & $-0.7479$ \\
Non-spin-flip, dipole
  & $1.4097$ & $-0.7997$ \\
\hline
\end{tabular}
\caption{Electric and magnetic response kernels for
$\Upsilon(1S)N$ scattering at $100\,\mathrm{MeV}$
above threshold and $Q^2=0.20\,\mathrm{GeV}^2$.
The matrix elements use fixed-axis canonical nucleon spinors in the
center-of-mass frame and the physical
quarkonium velocity. Errors on the cubic $z$ results
are propagated from the full lattice GFF covariance. The remaining rows show
the dependence on the form-factor
parametrization.}
\label{tab:physical}
\end{table}

The spin-flip channel displays a different interference pattern. At the same
kinematics,
$K_2^{\downarrow\uparrow}=0.0256\,\mathrm{GeV}$
and $K_S^{\downarrow\uparrow}=0.0433\,\mathrm{GeV}$.
The scalar and spin-two terms therefore partially
cancel in $K_E^{\downarrow\uparrow}$ and add in
$K_B^{\downarrow\uparrow}$. The magnetic spin-flip kernel is larger in
magnitude than the electric one. These are fixed-axis spin amplitudes, whose
individual signs and magnitudes depend on the spin convention. The scalar
spin-flip contribution reflects the overlap of the external
nucleon spinors in the chosen spin basis.

\begin{figure}[H]
\centering
\includegraphics[width=0.99\linewidth]{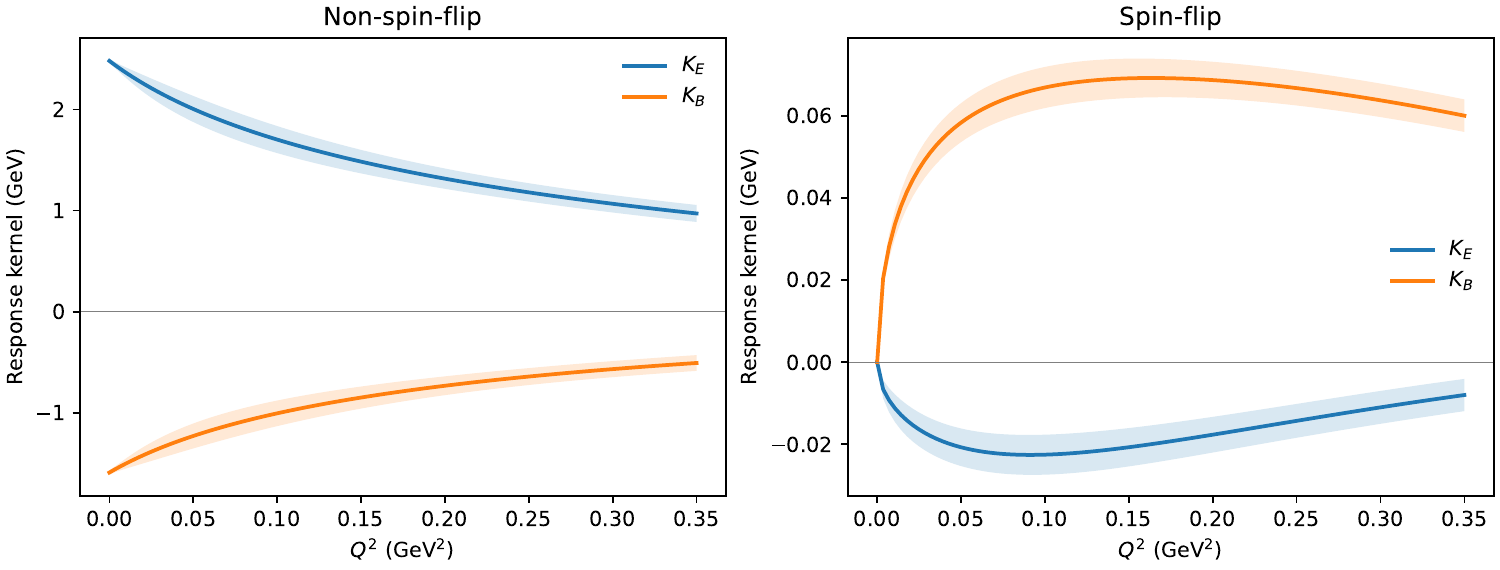}
\caption{Electric and magnetic $\Upsilon(1S)N$
response kernels as functions of momentum transfer
at $100\,\mathrm{MeV}$ above threshold.
The left panel shows non-spin-flip amplitudes and the right panel spin-flip
amplitudes, evaluated with fixed-axis nucleon spinors. The shaded
one-standard-deviation bands follow from the correlated lattice GFFs in the
cubic
$z$ fit, for the adopted scalar form-factor shapes.}
\label{fig:response}
\end{figure}

The relative importance of the electric and
magnetic contributions to the elastic amplitude
depends on both the nucleon response kernels and
the quarkonium polarizabilities, as expressed in
Eq.~\eqref{eq:electric-magnetic}.
Table~\ref{tab:physical} and Fig.~\ref{fig:response}
show the calculated kernels and their
momentum-transfer dependence.

The $0.1003\,\mathrm{GeV}$ standard deviation, or $7.62\%$ of the
central value, is propagated solely from the correlated lattice GFFs
for the specified chromoelectric non-spin-flip kernel. The corresponding
uncertainty of the full elastic amplitude also involves the quarkonium
polarizabilities, the scalar form factors, perturbative matching, and
corrections beyond the local interaction. The electric and magnetic
non-spin-flip kernels have a GFF-induced correlation coefficient of
$-0.976$, reflecting the opposite scalar contributions to the two
responses.

The scalar form factors and their momentum-transfer dependence introduce
further uncertainty. The sigma-term uncertainties contribute
$0.0176\,\mathrm{GeV}$ to the scalar-kernel
standard deviation when treated independently of the GFFs. Varying the light
scalar squared
radius from $1.34$ to $1.49\,\mathrm{fm}^2$
and the strange radius from $0$ to
$1.1\,\mathrm{fm}^2$ changes the electric
non-spin-flip kernel over a half-range
of $0.0236\,\mathrm{GeV}$.
At the selected kinematics, replacing the cubic
$z$ expansion with the dipole parametrization
shifts the electric non-spin-flip kernel by
$0.0934\,\mathrm{GeV}$, comparable to its
$0.1003\,\mathrm{GeV}$ uncertainty propagated
from the lattice GFF covariance.
This indicates a significant sensitivity to
the assumed form-factor parametrization.
The perturbative matching introduces an additional
dependence on the QCD scales. Varying the charm
matching scale from $1.10$ to $1.50\,\mathrm{GeV}$
changes the non-spin-flip electric kernel by at most
$0.14\%$ and the magnetic kernel by $0.10\%$ at the
selected energy and momentum transfer. The evolution
of the spin-two Wilson coefficients compensates that
of the quark and gluon GFFs, preserving their combined
contribution under changes of the renormalization scale
at leading-logarithmic accuracy. This residual threshold-scale
variation measures the sensitivity within the adopted matching
prescription. Higher-order scalar matching
and spin-two evolution contribute additional perturbative effects,
separate from the GFF covariance errors in Table~\ref{tab:physical}.

The contraction with the physical quarkonium
four-velocity determines the kinematic weighting
of the spin-two nucleon response.
To quantify its effect, we consider the
spin-averaged electric kernel norm,
\begin{equation}
\|K_E\|_{\rm spin}
=
\left[\frac12
\operatorname{tr}(K_EK_E^\dagger)\right]^{1/2}.
\label{eq:spin-norm}
\end{equation}
At the selected kinematics, this gives
$1.3164\,\mathrm{GeV}$, compared with
$1.2599\,\mathrm{GeV}$ when the physical
quarkonium velocity is replaced by the
normalized average nucleon four-momentum
defined in \ref{app:breit}.
The resulting $4.48\%$ difference reflects
the sensitivity of the leading local response
to the choice of contraction vector.

\subsection{Physical scope of the local response}
\label{subsec:eft-error}

Beyond the local interaction, spatial variations of the chromofields
and the propagation of color-excited heavy-quark configurations produce
additional momentum- and energy-dependent contributions.

At $Q^2=0.20\,\mathrm{GeV}^2$, the
three-momentum transfer in the center-of-mass
frame has magnitude $Q=0.447\,\mathrm{GeV}$.
The net energy transfer vanishes in this frame,
while the individual chromofield frequencies
in the two-field response are determined by
the nonlocal gluonic correlations in the nucleon.

The leading frequency dependence of the electric response can be related
to E1 transitions through intermediate color-excited states. Let
$\Delta_n=E_n-E_\Phi>0$ and
$\mathbf d_{n\Phi}=\langle n|\mathbf d|\Phi\rangle$, with the color
coupling included in the dipole operator. For an isotropic stable state,
and frequencies below the excitation spectrum, the crossing-even response
obeys
\begin{equation}
\begin{aligned}
\frac{\alpha_E(\omega)}{\alpha_E(0)}
&=1+\frac{\omega^2}{\Delta_{\rm eff}^2}+O(\omega^4),\\
\frac{1}{\Delta_{\rm eff}^2}
&=\frac{\displaystyle\sum_n
 |\mathbf d_{n\Phi}|^2/\Delta_n^3}
 {\displaystyle\sum_n
 |\mathbf d_{n\Phi}|^2/\Delta_n}.
\end{aligned}
\label{eq:nonlocal-response}
\end{equation}
The sums include the appropriate continuum of intermediate states.
The first frequency-dependent coefficient probes a higher inverse moment
of the excitation spectrum than the static polarizability. Its contribution
to quarkonium--nucleon scattering involves the corresponding nonlocal
chromofield correlations in the nucleon.

Spatial corrections to the leading dipole interaction
arise from higher chromoelectric multipoles and
covariant derivatives of the gluonic fields.
For an illustrative quarkonium separation
$r_\Phi=0.205\,\mathrm{fm}$, the selected momentum
transfer gives $(Qr_\Phi)^2=0.216$.
A characteristic soft-gluon momentum of
$0.3$--$0.5\,\mathrm{GeV}$ corresponds to
$(\Lambda_{\rm soft}r_\Phi)^2=0.10$--$0.27$.
The actual higher-multipole contributions depend
on the individual gluon momenta and the
corresponding quarkonium transition matrix elements.

For bottomonium, a characteristic internal velocity is
$v_b^2\sim0.1$ \cite{Trottier:1997}. Relativistic
corrections to the heavy-quark Hamiltonian and E1
transition operators arise at relative order $v_Q^2$
in pNRQCD, as established for radiative E1 transitions
\cite{BrambillaPietrulewiczVairo:2012}. The corresponding
corrections to the two-chromofield response involve
the modified transition matrix elements and propagation
of intermediate color-excited states. They enter at
the same relative velocity order as the Coulombic
magnetic contributions discussed in
Sec.~\ref{sec:matching}, subject to the assumed
scale hierarchy. Higher multipoles and nonlocal
responses depend additionally on gluonic momenta and excitation
energies. The local amplitude uses the separate static electric and
magnetic polarizabilities.

Nonperturbative hybrid excitations modify the
intermediate-state propagation governing the
chromoelectric response
\cite{LakhinaSwanson:2004}.
The size of the resulting nonlocal corrections
depends on the excitation spectrum and
the corresponding quarkonium transition matrix
elements, together with nonlocal gluonic
correlations in the nucleon.

\section{Conclusions}
\label{sec:summary}

The scalar and spin-two contributions to quarkonium--nucleon
scattering produce distinct chromoelectric and chromomagnetic
responses. At the near-threshold kinematics considered here,
the scalar contribution dominates the non-spin-flip kernels,
enhancing the electric response and partially cancelling
the magnetic one. The spin-flip amplitudes exhibit a
different interference pattern, reflecting the nucleon
spin dependence of the gravitational form factors and
their kinematic projections.

Correlated lattice form factors determine the momentum-transfer
dependence and GFF covariance uncertainties of the nucleon kernels.
The physical velocity controls their collision-energy dependence through
the spin-two projection. At the selected kinematics, the electric
non-spin-flip kernel is
$1.3163\pm0.1003_{\rm GFF}\,\mathrm{GeV}$.
The corresponding elastic scattering amplitude is obtained by weighting
the electric and magnetic kernels with the quarkonium polarizabilities;
its theoretical accuracy also depends on the scalar form factors,
perturbative matching, and higher-order quarkonium response.

\appendix

\section{Scalar matching at the charm threshold}
\label{app:scalar-matching}

At the charm threshold, the four-flavor
quark-mass operator
$\mathcal O_{m,4}=\sum_{q=u,d,s,c}m_q[\bar q q]$
is matched onto the three-flavor basis.
Using Eq.~\eqref{eq:charm-NLO}, the charm
contribution is expressed through
$\mathcal S_g$ in the three-flavor theory.
The trace identity then relates this operator
to $\Theta_3$ and
$\mathcal O_{m,3}=\sum_{q=u,d,s}m_q[\bar q q]$.

At the matching scale $\mu_c$, define
\begin{equation}
\begin{aligned}
a_3
&=\frac{
\beta^{(3)}(\alpha_s^{(3)}(\mu_c))
}{
\alpha_s^{(3)}(\mu_c)
},
\\
\kappa_c
&=-\frac{\alpha_s^{(4)}(\mu_c)}{3\pi}
\left[
1+\frac{11\alpha_s^{(4)}(\mu_c)}{4\pi}
\right].
\end{aligned}
\label{eq:scalar-threshold-definitions}
\end{equation}
The resulting scalar-coefficient matching is
\begin{equation}
\begin{aligned}
C_{\Theta,3}
&=C_{\Theta,4}
+\frac{\kappa_c}{a_3}
 C_{m,4}^{\mathrm{inv}},
\\
C_{m,3}^{\mathrm{inv}}
&=C_{m,4}^{\mathrm{inv}}
\left[
1-\frac{\kappa_c}{a_3}
(1+\gamma_m^{(3)})
\right].
\end{aligned}
\label{eq:scalar-threshold}
\end{equation}
The coupling constants and mass anomalous
dimension are evaluated at $\mu_c$ in their
respective flavor theories.

\section{Symmetric nucleon Breit projection}
\label{app:breit}

In the nucleon Breit frame, the momentum transfer
is purely spatial and the initial and final nucleon
three-momenta are opposite:  $\mathbf{P}'=-\mathbf{P}$.
The average momentum
$\bar P=(P'+P)/2$ is therefore
$\bar P^\mu=(E_B,\mathbf 0)$, with
$E_B=\sqrt{M^2-t/4}$. Its normalized four-vector is
$v_B^\mu=\bar P^\mu/E_B=(1,\mathbf 0)$.
For equal canonical-spin nucleon states,
contraction of Eq.~\eqref{eq:emttensor} gives
\begin{equation}
\begin{aligned}
\mathcal F_{i,B}^{(2)}(t,\mu)
&=\sqrt{1-\frac{t}{4M^2}}
\left[\frac{3M}{4}A_i(t,\mu)
+\frac{t}{16M}\bigl(3B_i(t,\mu)-D_i(t,\mu)\bigr)\right],\\
\mathcal A_{\Phi N,B}(t)
&=\sqrt{1-\frac{t}{4M^2}}
\left[C_\Theta^\Phi\Theta_N(t)
+C_m^{\Phi,\mathrm{inv}}\sigma_N(t)\right]
+\sum_{i=q,g}C_i^{(2),\Phi}\mathcal F_{i,B}^{(2)}.
\end{aligned}
\label{eq:breit}
\end{equation}
The Breit velocity contracts the EMT along the average nucleon momentum.
In the forward limit at zero relative velocity, this projection reproduces
the Luke--Manohar--Savage interaction in the corresponding normalization
\cite{LukeManoharSavage:1992}.
Writing $W=\sqrt{s}=M_\Phi+M+\epsilon$, the forward spin-two ratio becomes
\begin{equation}
\frac{\mathcal F_i^{(2)}(s,0)}{\mathcal F_{i,B}^{(2)}(0)}
=\frac{4\gamma_{\rm rel}^2-1}{3},\qquad
\gamma_{\rm rel}=\frac{s-M_\Phi^2-M^2}{2M_\Phi M}.
\label{eq:recoil-ratio}
\end{equation}

\section*{Acknowledgements}
The author thanks Professor Phiala E. Shanahan for generously providing the
correlated nucleon gravitational form factors and their covariance. The author
also thanks Dr. Yuxun Guo for kindly sharing lattice form-factor subsets and
form-factor samples used for comparison.

\section*{Declaration of competing interest}
The author declares that he has no known competing financial interests or
personal relationships that could have appeared to influence the work reported
in this paper.

\section*{Data availability}
The numerical results use correlated lattice nucleon form factors provided by
P. E. Shanahan. The underlying data were provided by the lattice collaboration.
The calculated quantities and numerical programs are available from the author
on reasonable request. Redistribution of the source data remains subject to the
providers' permission.

\section*{Declaration of generative AI and AI-assisted technologies in the
manuscript preparation process}
The author used OpenAI GPT-5.6 for manuscript organization, language editing,
literature review, and assistance with numerical plotting. The author takes
full responsibility for the scientific content, calculations, and conclusions.

\end{document}